\documentstyle[12pt,aasms4]{article}
\input epsf.tex
\begin{document}
\title{On the frequency and remnants of Hypernovae } 
\author{Brad M.S. Hansen }
\affil{Canadian Institute for Theoretical Astrophysics \\
University of Toronto, Toronto, ON M5S 3H8, Canada}
\authoremail{hansen@cita.utoronto.ca}
\begin{abstract}
Under the hypothesis that some fraction of massive stellar core collapses give
rise to unusually energetic events, termed hypernovae, I examine the required
rates assuming some fraction of such events yield gamma ray bursts. I then discuss evidence from
studies of pulsars and r-process nucleosynthesis that independently suggests the existence of
a class of unusually energetic events. Finally I describe a scenario which links
these different lines of evidence as supporting the hypernova hypothesis.
\end{abstract}


\section{Introduction}

Recent studies of Gamma-Ray Burst afterglows (van Paradijs et al 1997; Costa et al 1997; 
Frail et al 1997; Kulkarni et al 1998a, Galama et al 1998) and the determination of
some host or intervening galaxy redshifts (Metzger et al 1997, Kulkarni et al 1998b;
Djorgovski et al 1998) have indicated the presence of one or more new classes of
astrophysical explosion. The tentative identification of extragalactic star-formation
regions as the site of these events (Paczynski 1998; Djorgovski et al 1998) suggests a link
of such events with massive star evolution. In this letter I shall consider the possibility
that these events represent unusually energetic stellar explosions, 
 termed `hypernovae' in the literature.

Given a new class of astrophysical object or event, several natural questions arise.
What is the frequency with which it occurs (in our Galaxy and others)? What, if any,
are the observable manifestations of such an event in manners other than that of the
discovery?

\section{Rate Estimates}

Estimates of the cosmological Gamma-Ray Burst (or GRB) rate range from $\sim 10^{-6} {\rm yr^{-1}}$
per Galaxy for a constant comoving rate (Cohen \& Piran 1995) to $\sim 10^{-8} {\rm yr^{-1}}$
per Galaxy (Totani 1997; Wijers et al 1998) for a population that follows the cosmological
star formation rate (Lilly et al 1996; Madau et al 1996). These estimates assume an isotropically
emitting source. The estimated observed $\gamma$-ray energy of the GRB 971214
 is $\sim 3 \times 10^{53}$ ergs (Kulkarni
et al 1998b),  300 times the total energy output of an average supernova. This
may be reduced drastically if the emission is strongly beamed, but the corresponding
event rate increases by the same factor as the energy is reduced. Thus, one may scale
the event rate according to the true energy output of the average GRB,
\begin{equation}
R \sim 10^{-7} {\rm yr^{-1}} \left( \frac{3 \times 10^{52} ergs}{\epsilon E} \right)
\end{equation}
where I have used the star formation rate estimate (The comoving rate will be 100 times
larger). E is the total energy release and $\epsilon$ is the efficiency of conversion
to observed $\gamma$ ray energy.

Observations of a supernova 1998bw associated with the GRB980425 (Galama et al 1998;
Kulkarni et al 1998a) suggests that there exists a second class of
 GRB events which are indeed associated with the explosion
of a massive star and with $\gamma$-ray energy output  ($\sim 10^{48}$ ergs) significantly
less than the few other bursts with known distances. Thus, bursts
of this type can only be detected out to distances $\sim 100$ Mpc (Bloom et al 1998), as opposed
to the Gpc distances to other detected bursts. Hence the detectable volume for such events is
$\sim 10^{-3}$ that for the cosmological bursts and, given that as much as $\sim 10\%$ of observed bursts
could belong to this second class (Bloom et al 1998), the intrinsic comoving event rate is almost
certainly higher. 

 The association
of type Ib/c SN1998bw and GRB980425 have prompted some (Wang \& Wheeler 1998) to suggest an association
between GRB and all type Ib/c supernovae. This is disputed by several authors (Bloom et al 1998;
Kippen et al 1998; Graziani et al 1998) who point out the lack of other convincing associations as well as the 
unusually bright nature of SN1998bw. Although event rates based on a single event are necessarily uncertain,
I shall adopt a hypernova Ib rate $\sim 10\%$ that of the Supernova Ib rate. This is based on the fact
that Kippen et al present a catalogue of 160 bright supernovae (selection effects are claimed to be less
important for this subset)  since 1991 (the BATSE era), which contains 11 type Ib supernovae. The 10$\%$
hypernova fraction is high enough to allow the detection of at least one hypernova from samples of this
size while remaining consistent with the lack of other convincing associations (Bloom et al 1998; Kippen et al 1998;
Graziani et al 1998).
The Supernova Ib rate is approximately half that of type II supernovae (van den
Bergh \& Tammann 1991). Thus, I shall adopt a rate of $\sim 10^{-3} {\rm yr^{-1}}$  in the Galaxy as a hypernova rate.

\section{A Possible Class of Hypernova Remnants}

The offspring of supernovae are believed to be neutron stars (Baade \& Zwicky 1934). This
is supported by the association of some young pulsars with supernova remnants
and the approximate agreement of the pulsar birthrate with the supernova rate (Helfand
\& Becker 1984; Weiler \& Sramek 1988; Gaensler \& Johnston 1998). In some scenarios,
the offspring of a hypernova is an isolated black hole (Woosley 1993; Paczynski 1998),
which powers the GRB from either the binding energy of accreted material
or by magnetic field extraction of rotational energy. In the latter case, the rotational
energy required implies that the massive stellar core of the pre-collapse giant star
must be spinning rapidly with respect to the overlying envelope, contrary to some
evolutionary calculations (Spruit \& Phinney 1998 and references therein).
Others (Wang \& Wheeler 1998; Cen 1998)
have suggested that the high mean velocities of the pulsars (Lyne \& Lorimer 1994) result from
hypernova-like processes. However, these authors claim associations between all supernovae
of type Ib/c and GRB, which seems unlikely (Bloom et al 1998; Kippen et al 1998; Graziani et al 1998).

Here I suggest a modified version of the above scenario.
Recent work on the distribution of pulsar velocities, incorporating information from
different sources  such as pulsar-supernova remnant associations (Kaspi 1996), X-ray binary
properties (Brandt \& Podsiadlowski 1995; Kalogera, King \& Kolb 1998) as well as improved
treatments of the proper motion data (Hansen \& Phinney 1997; Hartman et al 1997;
 Cordes \& Chernoff 1998) all favour
a lower median velocity of $\sim 200-300 {\rm km.s^{-1}}$. However, there is dramatic evidence in some
individual cases for very high pulsar velocities (e.g. Cordes, Romani \& Chernoff 1993; Cordes \& Chernoff 1998). The
detailed statistical studies indicate that the fraction of pulsars with velocities $>800 {\rm km.s^{-1}}$
is less than 20\%(Hansen \& Phinney 1997; Cordes \& Chernoff 1998). This curious bimodal
distribution is, as yet, unexplained. 

If we associate the  pulsars in the lower velocity, majority component with the ordinary
supernovae (birthrate $\sim 10^{-2} {\rm yr^{-1}}$), then the high velocity pulsars represent a population with a birthrate appropriate
to that of the hypernovae $\sim 10^{-3} {\rm yr^{-1}}$. Furthermore, modelling of the SN1998bw
lightcurve (Iwamoto et al 1998; Woosley, Eastman \& Schmidt 1998) suggests an energy release
$\sim 3 \times 10^{52}$ ergs, approximately 30 times that of a traditional supernova. If we
regard this extra energy as the defining characteristic of a hypernova, we might expect a
consequently higher velocity for the remnant as well. Indeed, if the fraction of the total
energy channelled into pulsar kinetic energy is constant, the median velocity of pulsars
born from hypernovae is $\sim \sqrt{30} \times 200 {\rm km.s^{-1}} \sim 1100 {\rm km.s^{-1}}$, appropriate for the observed fast pulsar
population. However, the exact link between energy release and pulsar velocities is still
unknown.
 Proposed scenarios range from hydrodynamic or global instabilities and asymmetric collapse
(Burrows, Hayes \& Fryxell 1995; Janka \& M\"{u}ller 1996; Goldreich, Lai \& Sahrling 1998) to various
anisotropic radiation (`rocket') mechanisms (Harrison \& Tademaru 1975; Chugai 1984; Vilenkin 1995;
Kusenko \& Segre 1996; Horowitz \& Li 1997; Lai \& Qian 1998).

What of the spins of the hypernova pulsar offspring? Spruit \& Phinney (1998) have conjectured that the velocities
and spins of the observed pulsars may have the same origins (since an off-centre kick will generate
both linear and angular momentum). If that holds true in this case as well, we expect the spins
of the high velocity pulsars to also be particularly rapid. The initial spins of pulsars are
difficult to determine, since young pulsars undergo rapid spin down, but the fastest rotating
young X-ray pulsar rotates at a period of 16ms (Marshall et al 1998). This suggests that initial
spins $< 3$ms may be possible in hypernova events. I shall return to this point in the next section.

Hypernovae may also find application in the study of the production of r-process material. The most
widely accepted site of r-process production is the neutrino-heated ejecta of hot protoneutron stars
(Woosley \& Baron 1992; Meyer et al 1992). However, Qian, Vogel \& Wasserburg (1998) find that the
production of $^{129}$I and $^{180}$Hf for the protosolar nebula requires at least two different
production sites, with different ratios of neutrons to seed nuclei. They find that the two hypothetical
processes have to occur at different rates, with the less frequent events occurring at a rate
$\sim 1/10$ as often as the more frequent events and with lower ratios of neutrons to seed nuclei.
If we associate the high rate option with traditional core collapse supernovae, then our inferred hypernova
rate is appropriate to be the second kind of event. Furthermore, the r-process operates on neutrino
diffusion timescales $\sim$1-10s and on length scales corresponding to the neutrino-heated `hot bubble'
surrounding the nascent neutron star $\sim 10-50$km, where the mostly dissociated material yields
a high neutron/seed nuclei ratio (Meyer et al 1992). A neutron core moving at velocities $\sim 1000~{\rm km.s^{-1}}$ will cross this
bubble length within $\sim 0.1$ seconds. The neutrino-heated wind velocities on these
scales are $\sim 100-1000 {\rm km.s^{-1}}$ also (Qian \& Woosley 1996). Thus, the velocity of the neutron
star is likely to have a significant effect on the nucleosynthetic yield.
If hypernovae
 result in black hole remnants, they will not contribute to this process, as they swallow most
of their heavy element production (Timmes, Woosley \& Weaver 1996 and references therein).

Finally, it is worth noting that the distance ($\sim 40$~Mpc) of SN1998bw/GRB980425 is approximately
the value of the Greisen-Zatsepin-Kuz'min cutoff, estimated to be $\sim 30$~Mpc (Protheroe
\& Johnson 1995). Thus, if GRB events are responsible for the generation of Ultra-High Energy
cosmic rays (Milgrom \& Usov 1995; Waxman 1995), there are reasonable prospects for detection
of Cosmic Rays associated with this event. Recall that delays of $\sim$ 1 year are expected due to
Galactic magnetic fields.

\section{Magnetars and Cosmological Bursts}

If the spins of pulsars are determined by the kicks they receive during their birth, as suggested
by Spruit \& Phinney (1998), then the spins of pulsars born from hypernovae will be particularly
fast. Indeed, if spins reach $< 1$ms, the conditions for efficient field amplification by
proto-neutron star convection are met (Duncan \& Thompson 1992) and the remnant will most
likely be a magnetar,\footnote{
The existence of such objects has recently been demonstrated by Kouveliotou et al (1998).} or 
neutron star with magnetic field $\sim 10^{15}-10^{16}$~G.
 However, 
there is likely to be a distribution of spins and some normal field pulsars must result, since
many of the known high velocity pulsars have average magnetic field strengths.

If some fraction of hypernovae do yield magnetars, these events may power the cosmological
GRB as well, providing a common origin for the two observed classes. 
Several authors (Usov 1992; Fatuzzo \& Melia 1993; Thompson 1994; 
Blackman, Yi \& Field 1996) have discussed powering cosmological bursts using high field
neutron stars, although the usual scenario invokes accretion induced collapse of a  strongly
magnetic white dwarf. The scenario presented here plumbs a different energy source to power
the burst in that the rotational energy of the magnetar arises from the same mechanism that taps the
explosion to provide the kick velocity.

The estimated birthrate of magnetars in the Galaxy (see  Kouveliotou et al 1998 and references therein)
suggest a rate of similar order of magnitude to the hypernova rate. Thus, the fraction of hypernovae
that yield magnetars is $f_m \sim 0.1-1$. Assuming that one requires $P < 1$ms to generate
a magnetar (Duncan \& Thompson 1992), cosmological GRB should then tap an energy
reservoir $ E > 2 \times 10^{52}$ergs in this scenario. If we wish to match a rate $\sim f_m \times 10^{-3} {\rm yr^{-1}}$
 $\sim 10^{-4} {\rm yr^{-1}}$ with the rate in equation~(1), then we need only a total energy
in the beam $\sim 3 \times 10^{49}$ergs, corresponding to a beaming angle $\sim 3$~degrees and an
efficiency of conversion of rotational to beamed energy of $\sim 10^{-3}$. If we use the constant
comoving rate estimate, then the beaming angle is $\sim 30$~degrees and conversion efficiency
$\sim 0.1$. Thus, this scenario can easily generate sufficient events with appropriate energies
and beaming angles. If only a fraction of magnetar births generate GRB, then we require a greater
efficiency and larger beaming angle.
 Further constraints on the beaming are possible by studying the effects
of afterglows in other wavebands (Perna \& Loeb 1998). The high spins appropriate to the
magnetars may also help to explain the variation in durations between bursts via the competition
between gravitational and electromagnetic radiation (Blackman \& Yi 1998).

\section{Constraints and Predictions}

The scenarios I have described above invoke the release of energy in the core collapse of a massive
star to power them. As such, there is little dependance on the stellar envelope composition. Thus,
just as we believe type II and type Ib supernovae to correspond to core collapse of stars with and
without hydrogen envelopes respectively, we must expect hypernovae to occur in both hydrogen-rich
and hydrogen-poor forms. SN1998bw is believed to be a type Ib hypernova and we might ask whether
there exist any candidates for type II hypernovae? One possible candidate for such an event would be 
SN1979c (Branch et al 1981) which outshone
most other type II by 2-3 magnitudes. Furthermore, this was a supernova of type II-L, a class
which Gaskell (1992) claims is the hydrogen-rich equivalent of the type Ib events, in that
this subset seem to be much closer to standard `bombs' than the full, rather heterogeneous,
type II sample. Gaskell's
estimate for the fraction of unusually luminous type II-L is $\sim 4-8\%$, consistent with
our assumption that the overenergetic fraction of all core collapse explosions is $< 20\%$.

How many hypernovae generate GRB? Let us first consider the SN1998bw/GRB980425 class. Kulkarni et al (1998a)
find evidence in the radio emission for a relativistic shock preceding the main shock. It is
thought that this decelerating shock may have generated the gamma-rays at an earlier time by an as yet
poorly understood mechanism. To generate such a shock a significant amount of energy $> 10^{49}$~ergs must
have been coupled to the outer $\sim 10^{-5} M_{\odot}$ of the stellar envelope. As such it is likely
that this type of GRB will be associated only with type Ib hypernovae (by virtue of the
smaller envelope mass and steeper density gradient).

If we believe that this model can also explain  the more energetic cosmological
GRB, the scenario requires the beaming of energy from a young magnetar. Whether such events can occur
in type~II hypernovae will depend on whether the jet can penetrate the overlying hydrogen envelope
while still avoiding the baryon loading problem. If not, we do not expect a GRB associated with such
an event. However, the rotational energy $\sim 10^{52}$ergs released is still a substantial fraction
of the hypernova energy and may perhaps result in observable asymmetry in the explosion. Such events
should be detected in high-z supernova or direct optical transient searches.

If we consider the possibility that GRB may be associated with cosmological type~II hypernovae,
what are the chances of observing such an association? 
Let us consider the detectability of a bright supernova such as SN1979c in each of the well-studied cosmological
afterglow cases. I model the peak flux of this event as a diluted black body of effective temperature
$\sim 13 000$~K as inferred from the parameters presented in Schmidt, Kirshner \& Eastman (1992) and 
Cappellaro, Turatto \& Fernley (1995).
The  maximum brightness  may be compared to
the observed afterglow or host galaxy emission at the appropriate redshifted time of maximum light
($\sim$ 7 days in the rest frame for SN1979C).
In all three cases with redshift information (Metzger et al 1997;Kulkarni et al 1998a; Djorgovski et al 1998), the
peak R magnitude is larger than the afterglow or host magnitude at the appropriate time. Furthermore, the sensitivity
to extinction is large since the observed emission is from the rest-frame UV. The detectability of type Ib events
($\sim$ 1.5 magnitudes fainter at peak) is even harder. At limiting magnitudes $R \sim 25$, type II hypernovae
are detectable out to $z \sim 1$ even for reasonable extinctions, but may be dwarfed by the GRB afterglow itself.

The connection between high velocities and spins proposed by Spruit \& Phinney (1998) and
the connection between rapid spins and strong magnetic fields proposed by Duncan \& Thompson (1998)
naturally leads to a halo of  magnetars and neutron stars about our Galaxy and others. This is, in
fact, the GRB scenario proposed by Duncan \& Thompson which sought to explain GRB as magnetic
reconnection events in the Galactic magnetar halo. Although I now invoke their births as the
source of the GRB, it  is possible that there is a third class of GRB event waiting to be
discovered\footnote{Perhaps some of the bursts with no observable optical afterglow could
arise in this extended halo}.

\section{Conclusions}

In this paper I have presented circumstantial evidence from pulsar velocity and r-process nucleosynthesis studies
which support the existence of another class of astrophysical explosion besides the supernovae, and with
a rate and properties similar to that inferred for the hypernovae. Such links are highly speculative, but, given the
complexity of the theory underlying these phenomena, any suggestion or hint of corroborating evidence is
invaluable. Furthermore, the conditions that are likely to result in a hypernova are appropriate for the
production of magnetars, which could generate the cosmological GRB as well.

An important point to note here is that the connection between kicks and spins proposed by
Spruit \& Phinney provides a new source of rotational energy to power cosmological GRB.
 This scenario is essentially the inverse of that proposed by Cen (1998) or Wang \& Wheeler (1998),
in that we invoke the kick mechanism (whatever that may be) to provide the energy source of the 
burst (rather than a momentum imbalance in the burst jet emission to provide the kicks). It may also
serve to alleviate the problems associated with strong core-envelope coupling in collapsar progenitor models for GRB.

Note also that this model rests on an (as yet) unknown mechanism for generating $\sim 20-30$ times the
canonical $\sim 10^{51}$~ergs of mechanical energy in a core collapse explosion. Under this hypothesis, hypernovae should
occur in both hydrogen-rich and hydrogen-poor form, just as core-collapse supernovae do. However, it
may be more difficult to generate GRB if there is a massive hydrogen envelope to penetrate. Nevertheless,
both events should appear in optical transient searches that don't trigger on gamma rays.

 This model provides an explanation for the curious bimodality in the pulsar velocity distribution,
given the reasonable assumption that the contribution to kinetic energy is an approximately constant
fraction of the collapse energy release.
However, it must be noted that this is based on a sample of $\sim 100$ objects and pulsar surveys
are bedevilled by myriad selection effects. Ongoing observational programs will add to the data
in forthcoming years and should conclusively address the veracity of the velocity bimodality.

\clearpage

\end{document}